\documentclass{article}
\usepackage{emulateapj,graphics,onecolfloat,amsmath}
\usepackage{epsfig}

\input epsf
\lefthead{Schulz et. al.}
\righthead{Air Breakdown Modeling with ICEPIC}
\begin{document}

\twocolumn[
\title {Preliminary Modeling of Air Breakdown with the ICEPIC code}
\vspace{0.5 cm}
\author {A.E. Schulz$^{1,2}$, A.D. Greenwood$^{1}$, K.L. Cartwright$^{1}$,
P.J. Mardahl$^{1}$, R.E. Peterkin$^{1}$, N. Bruner$^{3}$, T. Genoni$^{3}$, 
T.P. Hughes$^{3}$, D. Welch$^{3}$}
\vspace{0.2 cm}
\affil{$^{1}$Air Force Research Laboratory, Directed Energy Directorate,
Kirtland AFB, NM 87117}
\affil{$^{2}$Department of Physics, Harvard University,
Cambridge, MA 02138}
\affil{$^{3}$Mission Research Corporation, Albuquerque, NM 87110}
\vspace{1 cm}
]

\rightskip=0pt
\section*{Abstract}
Interest in air breakdown phenomena has recently been re-kindled with 
the advent of advanced virtual prototyping of radio frequency (RF)
sources for use in 
high power microwave (HPM) weapons technology.  Air breakdown phenomena 
are of interest because the formation of a plasma layer at the aperture
of an RF source decreases the transmitted power to the target, and in 
some cases can cause significant reflection of RF radiation.  Understanding 
the mechanisms behind the formation of such plasma layers will aid in the 
development of maximally effective sources.  This paper begins with some of 
the basic theory behind air breakdown, and describes two independent approaches to 
modeling the formation of plasmas, the dielectric fluid model and the Particle
in Cell (PIC) approach.  Finally we present the results of preliminary studies
in numerical modeling and simulation of breakdown.
   
\section{Introduction}
The subject of RF breakdown of air is one of particular interest
to any current effort in designing a device 
to project a high power electromagnetic signal over some
distance through the atmosphere.  In the case of high power 
microwave devices, air breakdown interferes with
the transmission of RF power, and in some cases reflects a
large fraction of that power back toward the source.  It is 
useful to study the details of how the breakdown occurs in order
to answer several important questions. At what point after initial exposure 
to the RF signal does the breakdown occur?  What field 
intensities cause it to begin?  What densities are reached, and 
what is the density profile throughout the gas being ionized?  What
ionization states are produced?  How do the properties of the medium 
change, and what is the net effect on the propagation of the RF
signal? 

To begin answering these questions, it is necessary to investigate
the air chemistry and types of interactions that electrons
driven by the RF fields will have with the surrounding gas, and
also with
the device emitting the RF signal.  
High-energy electrons and ions that collide with background
neutral gas molecules can cause impact ionization. High-energy
electrons that collide with solid walls can produce secondary
electrons. Both types of collision events have the potential to
modify the plasma density, and hence can influence the physical
behavior of the device under investigation.  Thus, for more
realistic numerical simulations, we have begun to develop
numerical models for both of these time-dependent plasma formation
processes.   Including these effects 
will be crucial for any future virtual prototyping
of high power microwave devices, currently in development at the 
Air Force Research Laboratory.  Understanding how the properties of the
medium change will allow us to maximize the effectiveness of microwave
source design.

In this paper we begin a careful investigation of the
physics behind breakdown processes.  Section 2 outlines some basic 
theory surrounding the physics of breakdown.  Section 3 describes our
modeling techniques for propagating an electromagnetic signal through 
a medium, including a bulk treatment of the plasma as a dielectric fluid,
and also a model of plasma electrons using a Particle in Cell (PIC) 
approach.  Section 4 includes an investigation of the role of the 
collision frequency in the breakdown process.  Also in section 
4 we present the results of a full breakdown simulation in helium, a 
simpler gas than air, employing
the dielectric fluid breakdown model. Finally in section 5 we draw
conclusions and outline plans for future research.  

\vspace{1 cm}
\section{Breakdown basics}
The basic mechanism of RF breakdown is a catastrophic growth in
the density of electrons in the medium, driven by the energy taken
from the incident electromagnetic signal.  Several conditions are
necessary for this process to occur efficiently; there must be a
significant population of background electrons to begin the
ionization avalanche, there must be sufficient energy in the wave
to impart ionizing energies to the electrons being accelerated,
and the rate at which the background gas is being ionized must
exceed the rates at which electrons are lost through attachment,
diffusion and recombination.  The evolution of the number density
is governed by the equation
\begin{align}
{\partial N_e \over \partial t}=\lambda+\nu_i N_e-\nu_a N_e -\nu_d N_e -\zeta N_e^2
\end{align}
where the terms are: $\lambda$, the quiescent generation of electrons due
to background radiation such as cosmic rays, $\nu_i$, the ionization frequency, 
$\nu_a$, the attachment frequency, $\nu_d$, the diffusion frequency, and $\zeta$,
the recombination rate constant.
At atmospheric temperatures and
pressures $\nu_a >> \nu_d$ and $\nu_a >> \zeta N_e$, so attachment is
the loss mechanism which imposes the strongest constraint on the
breakdown threshold.  The ionization and attachment frequencies
are given by
\begin{align}
\nu_i = {\alpha e E_{eff} \over m \nu_c} \\
\nu_a = {\beta e E_{eff} \over m \nu_c}
\end{align}
where \begin{align} E_{eff}^2={E_{rms}^2 \over 1+\omega^2 /
\nu_c^2}
\end{align}
Here $\nu_c$ is the collision frequency that includes all momentum
transferring events, and $\alpha$, $\beta$, and $\nu_c$ all depend
on the mean electron energy $W$, which evolves according to the
equation
\begin{align}
{dW \over dt}={e^2 E{eff}^2 \over m \nu_c}-{2m \over M}\nu_c W
-\nu_i W
\end{align}
The first term in this equation represents the energy gained under
the influence of the RF field, and the loss terms are from
inelastic collisions and ionizing collisions respectively.
Functional forms for $\alpha$ and $\beta$ and much of the theory 
presented in this section can also be found in \cite{anderlisak}.

Just how the concept breakdown is defined is a matter of some
debate.  The conventional wisdom is that when the electron density
becomes so large that the plasma frequency grows comparable the
the frequency of the RF radiation, the signal can no longer
propagate through the medium, and breakdown has occurred.  We
agree that interference with the propagation of the signal is an
adequate definition of breakdown.  However we shall demonstrate
later in this section that because the electrons suffer collisions
with the background gas molecules, the threshold at which the
propagation is impacted can sometimes correspond to plasma
frequencies that are considerably higher than the frequency of the
RF signal.  Nonetheless, for the purposes of developing some
intuition about breakdown it is useful to compare the plasma
frequency with the RF frequency.

The extent of the breakdown that occurs depends somewhat on
whether the medium is exposed to a continuous or a pulsed RF
signal.  The breakdown criterion for a continuous signal is simply
that $\partial n/ \partial t>0$.  Because the signal never shuts
off, there is no time limit and the plasma frequency will
eventually grow to be comparable to the continuous wave frequency,
and thus affect the signal.  The only condition for breakdown via
continuous RF
radiation is that the energy be large enough to ionize more
electrons than are being lost to attachment, diffusion, and
recombination.

The breakdown criterion for a single pulse is a little more complex but
not difficult to understand.  Because of the limited pulse length,
$\tau$, $dn/dt$ must be large enough to make $\omega_p \sim
\omega_{RF}$ on timescales less than $\tau$.  If this condition is
not met, the pulse will pass through the medium unaffected. The
rate of density growth depends on the mean electron energy $W$,
which depends on the field strength.  Therefore, shorter pulses
will require higher field strengths than longer ones in order to
suffer breakdown effects.  When breakdown occurs in a pulse, the
effect is typically that the back portion is either absorbed or
reflected, which is often referred to as tail erosion.  Fig. (1)
depicts the passage of a square pulse through a region of gas
which in time is getting ionized and affecting the pulse shape. At
early times, the gas has a small ionization fraction and is
virtually transparent to the incident electromagnetic signal,
which explains why the front portion of the pulse is unaffected.
The radiation accelerates the few existing electrons, and begins
to lose its energy to heating and ionization of the background gas
molecules.  As the electron density grows, the plasma begins to
reflect an increasing fraction of the incident wave, resulting in
smaller transmitted power.  This is responsible for eroding the 
back half of the pulse.

\begin{figure}[h]
\centerline{\epsfysize=6.0cm \epsfbox{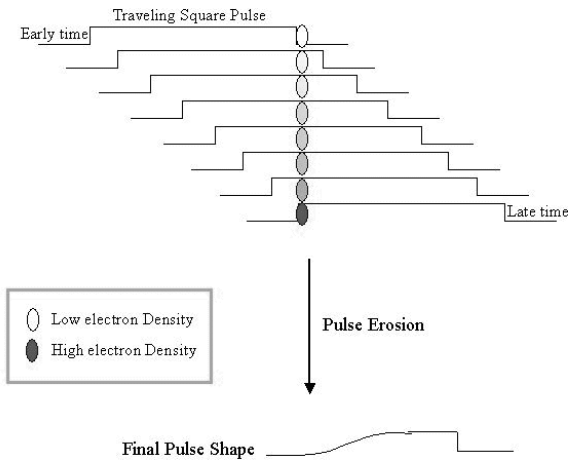}}
\caption{This cartoon shows how the passage of a pulse through a region of
gas causes ionization of the gas and an increase in the electron density.  As
the electron density grows, more of the energy in the pulse is being absorbed
or reflected, changing the shape of the wave-form that is transmitted through
the region.}
\end{figure}

The breakdown criterion for repeated pulses is similar to that of
the single pulse, with the added complication that while the
signal is off, the electron density is decreasing because of the
attachment, diffusion, and recombination loss mechanisms. Multiple
pulses relax the requirement on $\partial N_e / \partial t$ since
the driving field will return several times to feed energy into
the plasma.  The breakdown requirement therefore depends both on
pulse length $\tau$ and on the pulse repetition frequency
$\tau_{prf}$.  During the time when the pulse is off $\tau_{prf} -
\tau$, the electron density is decaying, so the breakdown
criterion for and infinite string of pulses will be that $\Delta
N_e({\rm on})>\Delta N_e({\rm off})$.

Introducing collisions into the picture complicates things
substantially.  The effect of a high collision frequency $\nu_c$
is to retard the rate of density growth, and also to retard the
erosion of the transmitted signal.  To understand this, consider
first that a high collision frequency means that the mean free
path and the mean free time between collisions are both very
short. On effect of the short mean free time is that the electron
being accelerated in the RF field does not often have time to
gather sufficient energy to ionize a gas molecule before it
collides again.  Each time it collides, it loses whatever small
energy it gained in the acceleration to the neutral atom it collided
with, so ionizing events become much rarer.  For the same reason,
the electrons are also much less efficient at removing energy from
the RF signal and dumping it into heating of the gas.

A consequence of the short mean free path resulting from a high
collision frequency is that the electrons become much less mobile.
Confining the electrons close to their original locations
effectively makes the plasma behave less like a metal than when
the electrons are highly mobile.  The frequent collisions prevent
the electrons from displacing from their ion friends and
counteracting the incident field.  Thus, a high collision
frequency greatly decreases the efficiency of reflecting the
signal, even when the electron densities are so high that
$\omega_p \sim \omega_{RF}$.  Fig. (2) illustrates this point by
showing the net transmitted signal through a plasma layer as a
function of the collision frequency with the background gas. The
plasma is held at a constant density, corresponding to a plasma
frequency of $\omega_p=21.45$.  On the left side of the plot the
transmission coefficients asymptote to their values in the
collisionless regime.  Four of the frequencies plotted are above
the plasma frequency, whereas the fifth is below it, and in the
collisionless case would be perfectly reflected by the plasma
layer.  In the case of the four transmitting frequencies, when the
collision frequency is increased (corresponding perhaps to a
higher density of background gas) the signal is being absorbed
into heating of the background gas.  Notice however that as the
collision frequency is increased further, the mobility and energy
transfer of the electrons is so hampered that all the frequencies
begin to transmit through the plasma layer, regardless of whether
$\omega_p$ is less than or greater than $\omega_{RF}$.

\begin{figure}[h]
\centerline{\epsfysize=6.0cm \epsfbox{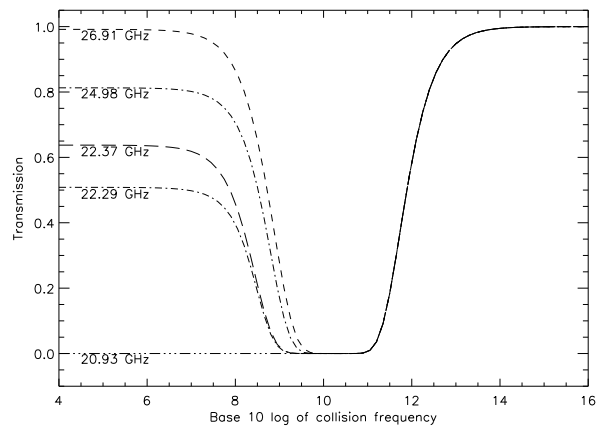}}
\caption{The theoretical prediction for the transmission coefficient for several
different RF frequencies as a function of the collision frequency $\nu_c$ of electrons
in the plasma with background neutral gas.  The
plasma density is held fixed, with a plasma frequency of $\omega_p=21.45$ GHz.}
\end{figure}

\vspace{1 cm}
\section{Method}
\subsection{The Dielectric Fluid Model}
For dense plasmas, it is a good approximation to treat
the plasma as a fluid with a frequency dependent dielectric constant.
In the presence of EM
radiation, this approximation is good for time-scales over which the
fluid moves a small amount.  The dielectric constant used in the
fluid approximation is (in terms of the permittivity of free space
$\epsilon_0$, the collision frequency $\nu_c$, and the frequency of the
electromagnetic radiation $\omega_{\text{\tiny EM}}$)
\begin{equation}
\epsilon=\epsilon_0 \left( 1-{\omega_p^2 \over \omega_{\text{\tiny EM}}
(\omega_{\text{\tiny EM}}
+ i \nu_c )} \right)
\end{equation}
where the plasma frequency $\omega_p$ depends on the density $n_0$
and species ($m$, $q$) of the ionized particles, and is given by
$\omega_p^2=n_0 q^2 / m \epsilon_0$.

To model high density plasma in the time domain, consider Ampere's law with
the plasma dielectric constant;
\begin{align}
\nabla \times {\bf B}({\bf x},\omega) &= \nonumber
\\ &\mu_0 \epsilon_0
\left( -i \omega +
{\omega_p^2 \over \nu_c - i \omega} \right) {\bf E}({\bf x},\omega)
\end{align}
The equivalent expression in the time
domain is given by
\begin{align}
\nabla\times{\bf B}({\bf x},t)&=
\mu_0 \epsilon_0 {\partial {\bf E}({\bf x},t) \over \partial t}
\nonumber \\
&+\omega_p^2 \mu_0 \epsilon_0 \int_{-
\infty}^t e^{- \nu_c (t-\tau)}{\bf E}({\bf x},\tau) d\tau
\end{align}
Differentiating Eq. (8) with respect to
time and substituting the result back into Eq. (8) eliminates
the convolution integral and yields the
following expression used to define the field update equations in the
dielectric fluid model.
\begin{align}
{\partial^2 {\bf E}({\bf x},t) \over \partial t^2}&+\nu_c {\partial {\bf E}(
{\bf x},t) \over \partial t}
+\omega_p^2 {\bf E}({\bf x},t) \nonumber \\
=&{1 \over \mu_0 \epsilon_0}
\nabla \times \left( {\partial {\bf B}({\bf x},t) \over \partial t} + \nu_c
{\bf B}({\bf x},t) \right)
\end{align}
This expression is used together with
Faraday's law used for updating ${\bf B}$
to define the field update equations in the
dielectric fluid model.
It can be shown that this method exhibits 2nd order
accuracy.  The demonstration of this accuracy and the details of the 
approximations used in this model can be found in \cite{us1}.

\subsection{Collisions and air chemistry in the fluid model}

   Breakdown models for helium, dry air, air with water content, argon and
SF$_6$ have been developed for the ICEPIC code.  These models are valid
for atmospheric pressures. 
The static parameters used in the model are the composition of the gaseous medium,
the gas temperature, and the gas density.  Several rates have been tabulated in
advance so that ICEPIC simulations need only reference look-up tables, instead
of numerically solving the Boltzmann equation on the spot.  As a function of
the magnitude of the RF field over the pressure $E/p$, the look-up tables
store the collision frequency, the ionization rate, the electron attachment
rates, recombination rates,
and the electron temperature.  The values in these tables are determined
by using the EEDF software package \cite{EEDF} to calculate the electron energy
distribution function in the gas mixtures.  This is done by using experimentally
determined interaction cross sections to numerically solve
the steady state Boltzmann equation for the isotropic part of the distribution
function.

The dry air model consists of a
mixture of $N_2$, $O_2$, and $Ar$ in a ratio of 78/21/1.
Several different air chemistry processes are modeled in the
fluid breakdown treatment.  They are outlined as follows:
\\ \\
\begin{tabular*}{5 cm}{l}
Elastic Collisions \\
$e^{-}+ M \rightarrow M+e^{-}$ \\ \\
Excitation \\
$e^{-}+ M \rightarrow M^{*}+e^{-}$ \\ \\
Ionization \\
$e^{-}+N_2 \rightarrow N_2^+ +2e^{-}$ \\
$e^{-}+O_2 \rightarrow O_2^+ +2e^{-}$ \\
$e^{-}+O_2 \rightarrow O^+ + O +2e^{-}$ \\
$e^{-}+Ar \rightarrow Ar^+ +2e^{-}$ \\ \\
3-Body Attachment \\
$e^{-}+O_2 + M \rightarrow O_2^{-} + M$ \\ \\
Dissociative Attachment \\
$e^{-} + O_2 \rightarrow O^{-} + O$ \\ \\
Recombination \\
$e^{-} + N_2^{+} \rightarrow N+N$ \\
$e^{-} + O_2^{+} \rightarrow O+O$
\end{tabular*}
\\ \\

   In the ICEPIC simulation, the electron density, collision frequency, and electron
temperature are all stored for each cell.  The lookup tables are used for all the gas mixtures
to evolve the number
density of electrons in each timestep according Eq. (1).
From this number density new effective values
of $\omega_p$ and $\nu_c$ are computed.  These quantities are used
to update the dielectric constant from Eq. (1).  Finally the ${\bf
E}$ and ${\bf B}$ fields are advanced in ICEPIC, and the process
begins anew.

The air model that
includes humidity is tabulated for four possible values of water molecule
content; 1\%, 2\%, 3\%,and 4\%, where 4\% is fully saturated air at room temperature.
Since the data are reasonably smooth functions, it is reasonable to interpolate
for intermediate values of $E/p$ and water content.  Water vapor is electronegative, so
the effect of introducing it into the gas mixture is to increase the attachment rates.
The ionization rate, however, is unaffected by the presence of water vapor. As a result,
the breakdown voltage increases with humidity.  One effect that has not been implemented
yet in this model, however, is the potential presence of water droplets in a humid
environment.  Because water droplets can create local field enhancements, they can in
principle increase the ionization rate.  Whether the increase in $N_e$ due to water
droplets is balanced by the decrease due to higher levels of humidity is a matter of
some debate, and will be further investigated in future work.

\subsection{The Particle In Cell (PIC) treatment}
ICEPIC computes the time advance of the
magnetic field according to Faraday's law, and the electric field according
to Ampere-Maxwell's law.  The discreet form of these equations used in
ICEPIC preserve the constraint
equations $\nabla \cdot {\bf B} = 0$
and $\nabla \cdot {\bf E} = \rho/ \epsilon_0$ as long as the initial data
satisfies these constraints.
The particles used in ICEPIC are ``macro-particles''
that represent many charged particles (electrons and/or ions) with a
position vector ${\bf x}$ and a velocity vector ${\bf v}=d{\bf x}/dt$.  The
relativistic form of Lorentz's force equation is used to determine the
particle's velocity:
\begin{equation}
{\bf F} = m{d \gamma {\bf v} \over dt} = q\left({\bf E} +{{\bf v} \over c}
\times {\bf B} \right)
\end{equation}
where $\gamma$ is the the usual relativistic factor of $(1-v^2/c^2)^{-1/2}$,
and $q$ and $m$ are the charge and mass of the particle.

ICEPIC uses a fixed, Cartesian,
logical grid to difference the electric and magnetic
field equations.
The vector quantities ${\bf E}$, ${\bf B}$, and ${\bf J}$
are staggered in their grid location using the technique of \cite{Yee}.
${\bf E}$ and ${\bf J}$ are located on the edges of the primary grid, whereas
${\bf B}$ is located on the faces of the primary grid.  An explicit leap-frog
time step technique is used to advance the electric and magnetic fields
forward in time.  The advantages of the leap-frog method are simplicity and
second-order accuracy.  The electric field advances on whole integer time
steps whereas the magnetic field and the current density advance on half
integer time steps.

The three components of the momentum and position of each particle are updated
via Eq. (10) using the Boris relativistic particle push algorithm
\cite{Boris}. The particle equations for velocity and position are also
advanced with a leap-frog technique.  The velocity components are advanced
on half integer time steps, and the particle positions are
updated on integer time steps.  The current density weighting employs an
exact charge conserving current weighting algorithm by \cite{Villa}, enforcing
$\nabla \cdot {\bf E} = \rho / \epsilon_0$.  Once the particles' positions
and velocities are updated and the new current density is updated on the grid,
the solution process starts over again by solving the field equations.

\subsection{Collisions and air chemistry with PIC}

The interactions suffered by PIC particles that are
being used to model air chemistry are electron-neutral scattering, excitation,
and ionization, and ion-neutral scattering, ionization, and charge
exchange.
Energetic particles that interact with a background gas of neutral atoms
have a probability of collision $P_i$ during a time interval $\Delta t$ that
depends on the number density of background neutral gas molecules $n_g$, the
energy-dependent cross-section $\sigma(E_i)$, and velocity $v_i$ through the 
collision frequency of the $i^{th}$ particle $\nu_i=n_g(x)\sigma(E_i)v_i$.
\begin{align}
P_i=1-e^{-\nu_i \Delta t}
\end{align}
One scheme for determining whether the $i^{th}$ particle collides and interacts in a
given timestep is to calculate
$P_i$ and compare it with a uniform
random number $R$. For $P_i>R$, the particle will be collided in this time
step.  However, determining collisions in this way can be computationally
expensive.  It is substantially less computationally expensive to use the
null-collision method in which we compute an energy independent collision
frequency:
\begin{align}
\nu_{null}={\rm MAX}_x(n_g(x)){\rm MAX}_E(\sigma(E)v)
\end{align}
In this approximation we need not calculate the cross section for every particle in
every timestep.  Rather, we construct a total collision probability $P_T$ that
represents the fraction of the particles that undergo a collision in a
single timestep:
\begin{align}
P_T=1-e^{-\nu_{null}\Delta t}
\end{align}
For multiple reactions with the same background gas, the cross-section
used in Eq. (12) is a sum of all the individual cross-sections.
This method is applied
for each background gas in the simulation.

The subset of the total number of particles undergoing a collision in a
time step is chosen randomly from the whole set.  From this fraction,
the energy-dependent cross-section is used to determine if a real
collision occurs.  Hence, if
\begin{align}
{n_g(x)\sigma(E_i)\nu_i \over \nu_{null}}> R
\end{align}
then a collision occurs.  For multiple reactions with the same background
gas, the particle must be tested to see if it under went a reaction with
each cross-section that was used to find the cross-section in Eq. (12).
Even if the particle undergoes a reaction, it still may experience a different
reaction in the same timestep, because it represents many particles.

The ion-neutral collisions are implemented using the same method. For
high-energy ions (100kV-500kV), the ion collision frequency is higher
than the electron collision frequency.

We have also developed a model based on previous work by Vaughan, Shih, and
Gopinath \cite{vaughan} for the secondary electron emission yield $\delta$.  This 
model is not used in the breakdown studies presented in this paper, but 
we mention it because it is a critical part of modeling any experiment
that has a vacuum window, such as the bell jar experiments currently 
being conducted that the Air Force Research Laboratory. The secondary 
electron emission yield depends on the
energy and angle-of-incidence of the primary electron,
\begin{align}
\delta&(E,\Theta)=\delta_{max 0} \left( 1+k_s{\Theta^2 \over 2 \pi} \right) \times
f(w,k)
\end{align}
where
\begin{align}
f&(w,k)=  \\ &\left\{
\begin{array}{lcc}
(we^{(1-w)})^k & k=\left\{\begin{array}{c} 0.62 \\ 0.25
\end{array} \right . &
\begin{array}{c} w\leq1\\ 1<w\leq 3 \end{array} \\
3.0w^{-0.35} && w>3
\end{array}
\right .
\nonumber
\end{align}
where the energy dependence appears implicitly via
\begin{align}
w={E-E_0 \over E_{max 0} \left( 1+k_s{\Theta^2 \over 2 \pi} \right)-E_0}
\end{align}

\vspace{1 cm}
\section{Preliminary results}
To date we have used both the particle and the fluid breakdown
representations to probe transmission through a plasma layer. We
have used both implementations to test our hypotheses regarding
the functional dependence of the transmission coefficient on the
value of the collision frequency. We have also used the fluid
representation to simulate the breakdown of a gaseous medium in a
two dimensional box.  We have observed the growth of the number
density of electrons in time and space, and have run the
simulation until a steady state is achieved.  We have not been
able to repeat this portion of the investigation with the PIC
representation, because of complications regarding resolution of
the Debye length, which we shall summarize along with proposed
solutions to the problem.

\subsection{The impact of collisions on transmission}
To test our hypotheses regarding the dependence of
transmission on the collision frequency, summarized in Fig. (2),
we have fixed the electron density in the breakdown models. Our
results are shown in Fig. (3).  It is worth mentioning that
$\omega_p$ in general depends on $\nu_c$ because $\nu_c$ enters
into the evolution of the number density of electrons.  However,
since we have fixed $\omega_p$ in this study, Fig. (3) reflects
only the impact of $\nu_c$ on the mobility and mean free time of
the plasma electrons, but not it's changes to $\omega_p$.  It
gives us intuition for how changing the mobility of the electrons
affects their efficiency at transferring the RF energy to the
background gas (lower values of $\nu_c$), and also how limited
mobility impedes their ability to reflect RF energy (higher values
of $\nu_c$), but it tells us nothing about how $\nu_c$'s changing
$\omega_p$ impacts the fraction of RF energy reflected.

\begin{figure}[h]
\centerline{\epsfysize=6.0cm \epsfbox{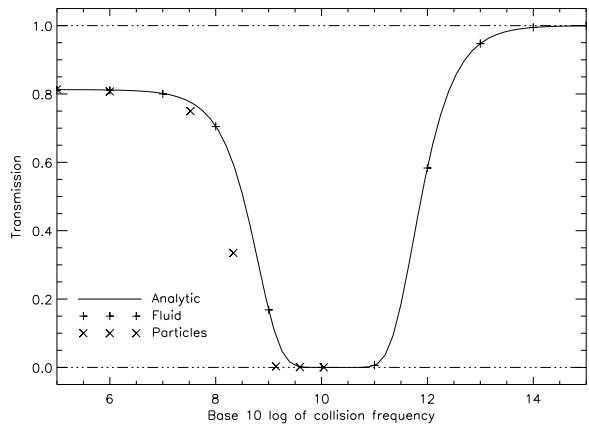}}
\caption{The transmission coefficient of an RF signal with $\omega=24.98$ GHz
as a function of the collision frequency with background neutrals
for a fixed plasma density corresponding to $\omega_p=21.45$ GHz.  Particle
collision frequencies have been corrected for grid heating effects.}
\end{figure}

In the case of the dielectric fluid model, fixing the density is
an easy modification, we simply ran the simulations without
evolving the number density and keeping the plasma frequency fixed
at a particular value, but varying the collision frequency in the
dielectric constant from Eq. (1).  In the PIC case we have
artificially enforced that the number density remain the same by
allowing momentum scattering collisions to take place, but not adding
another particle to the simulation.  Changes to the collision
frequency were accomplished by making the corresponding change to
the interaction cross section. The results from the PIC
computation require a more sophisticated interpretation, however,
because in these runs we had insufficient computational resources
to adequately resolve the Debye length. The effect of this lack of
resolution of the Debye shielding is that particles in the
simulation experience electromagnetic forces from other particles
that ought to have been shielded, which causes all the particles
in the simulation to heat.   The collision frequency in the PIC
representations depends on the temperature as well as on the cross
section $\sigma$, according to the formula
\begin{align}
\nu_c={m_e \sigma^2\over 3 k_B T_e n_g}
\end{align}
Therefore, a correction for the particle heating was introduced in
order to properly plot transmission versus collision frequency.
The correction we applied was a very rough one, essentially taking
the average temperature of all the particles in the simulation at
the very end of the run, and using that temperature along with the
cross section we supplied to calculate $\nu_c$, the independent
variable in Fig.(4).

Another difficulty with the particle simulations occurs when the
collision frequency grows enough to require multiple collisions
per cell, per timestep of the simulation.  This essentially causes
the run time to increase dramatically, and we did not complete any
simulations for collisions frequencies past this threshold.  In
future it would be extremely useful to probe this part of the
parameter space with the particle collision model, therefore the
development of a more efficient treatment of particle collisions
is a matter of current investigation.

\subsection{Breakdown simulation in a 2 dimensional box}
In a  second study we conducted a full breakdown simulation of an
RF signal incident upon a region of helium gas in a 2-D box. 
We chose to study helium because there are fewer interaction 
processes and thus it was one of the first to be tabulated 
and implemented in ICEPIC.  The interactions modeled in the helium
breakdown are
\\ \\
\begin{tabular*}{5 cm}{l}
Elastic Collisions \\
$e^{-}+ He \rightarrow He+e^{-}$ \\ \\
Excitation \\
$e^{-}+ He \rightarrow He^{*}+e^{-}$ \\ \\
Ionization \\
$e^{-}+He_2 \rightarrow He_2^+ +2e^{-}$ \\ \\
Recombination \\
$e^{-} + He_2^{+} \rightarrow He^{*}+He$ \\ \\
\end{tabular*}
\\ \\ 
The geometry used in the simulation is shown in Figure (4).  The probe 
recording the transmitted signal is placed in the right vacuum outside the 
helium layer, but within the plane wave launching boundary.  The probe 
measuring the reflected signal is in the left vacuum outside the plane wave
launcher, to avoid interference with the incident wave.  
\begin{figure}[b]
\centerline{\epsfysize=3.9cm \epsfbox{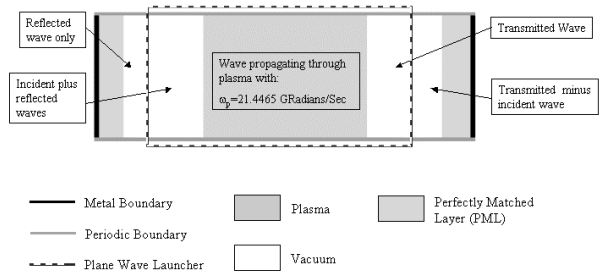}}
\caption{A helium layer lies between two vacuum regions.  The dashed line 
represents the plane wave launching boundary, which emits on the left and 
absorbs on the right.  The top and bottom boundaries of the simulation are periodic,
while the left and right boundaries are metallic but shielded with a Perfectly
Matched Layer (PML) that attenuates all EM signals to 30 db.}
\end{figure}

To study the breakdown, we made movies of the $|{\bf E}|$ field in the box, 
and observed the process as the RF energy  was first attenuated in the helium, 
and later reflected, and finally reached a steady state with some fraction
being transmitted through, and a larger fraction being reflected back toward the
source.  We also made a simultaneous
movie of the density of free electrons in the plasma, which was very helpful in
understanding the various stages in the $|{\bf E}|$ field movie.   

In the beginning of the simulation, the RF wave hits the helium layer and 
initially passes directly through it virtually 
unaffected.  In the electron density 
movie however, the electron density begins to grow at 
a constant rate throughout the medium.  The electron density is higher at the 
left edge 
of the helium layer than further in because more of the wave causing the ionization
has passes by that point.  After some time, the amplitude of the RF signal 
begins to get attenuated, and simultaneously in the density movie, the density
of electrons near the vacuum-helium interface begins to grow exponentially.  This
attenuates the wave enough that the density further into the helium layer stops 
growing and in some cases, may even decrease.  This happens because the signal
that was driving the ionization of the background gas is no longer strong enough
to do so.  From this time forward, any fraction of the RF signal that makes it 
through the high density region at the left interface with the vacuum propagates without
any significant
attenuation through the rest of the helium layer.  The action at the interface
has just begun, however.  As the electron density grows exponentially, the surface
of the plasma becomes much more efficient at reflecting the signal, and thus the 
${\bf E}$ field movies shows a growth in the reflected power and a corresponding 
decrease in the net transmitted signal.  This causes the exponential growth of 
the electron density near the boundary to cease, and eventually a stable equilibrium
is reached with a static distribution in the electron density, and a fixed ratio
between the transmitted and reflected signals.  Figs. (5) and (6) 
are frames from the ${\bf E}$ field and electron density movies, and show 
the long term steady state equilibrium that is reached in the field values and 
density distribution.  

\begin{figure}[h]
\centerline{\epsfysize=6.0cm \epsfbox{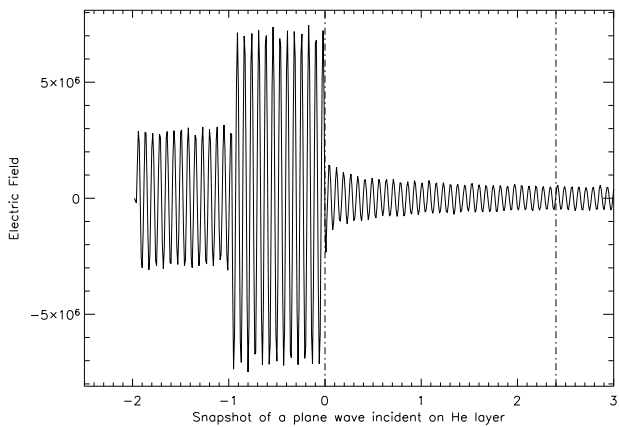}}
\caption{This is a snapshot of the traveling E-M wave though the helium layer.  The
helium is located between 0 and 2.4 on the plot.  The wave emitter is located at 
the position -1.  The signal between -2 and -1 is traveling from right to left, 
and represents the net reflected signal from the helium layer.  The wave between
-1 and 0 is the interference between the incident and reflected waves, currently
undergoing constructive interference. The transmitted signal is measured outside
the helium, in the region between 2.4 and 3.  }
\end{figure}

\begin{figure}[h]
\centerline{\epsfysize=6.0cm \epsfbox{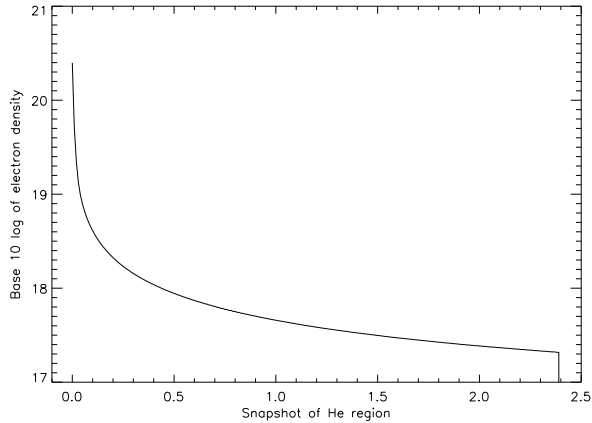}}
\caption{This is a snapshot of the density profile of the free electrons in the plasma.
This illustrates why most of the RF signal is reflected or attenuated between 0 and 0.2.
The initial density profile was flat at $0.5 \times 10^{16}$ throughout the helium.  The 
RF signal drove a constant growth in density until the exponential growth at the left interface 
prevented sufficient transmission to continue ionizing the plasma further in.}

\end{figure}

It would have been a nice experiment
to repeat this analysis using the PIC representation of
breakdown physics, and compare the predictions of each model quantitatively. 
This has turned out to be a serious challenge for the following reasons.  First, 
by examining the density plot in Fig (6) it is clear that the number of 
particles has grown over several orders of magnitude since the simulation began.
This presents a problem for PIC not only because there are so many particles, but
because these high densities require very fine resolution in the mesh to adequately
resolve the Debye length.  Simulations run without resolution of the Debye Length at the
densest regions are subject to severe grid heating of the particles, which in turn 
affects the collision frequency with the background gas, and in general causes 
incorrect predictions in the physics.  Even though the physical area that requires this
high level of resolution is only a small fraction of the total simulation area, 
ICEPIC does not yet have adaptive mesh capability, and would thus require that 
we resolve the entire domain on this small scale.  Another difficulty is that the collision 
frequency is growing substantially throughout the simulation.  When the collision frequency
becomes so high that random numbers must be generated many times for each cell 
in each timestep, it causes the simulation to grind to a near halt.  However, there is yet
hope.  The breakdown simulation performed with the bulk model suggests that the majority 
of the interesting physical phenomena take place within a short distance of the interface
with the vacuum.  Thus, it is not necessary to simulate such a vastly large gaseous layer, 
and we may be able to compare direct results between the PIC and fluid breakdown
models if we model transmission through a very thin region.  To do this successfully however
will still require that we develop a better way to model multiple collisions per timestep in
the PIC regime.  These subjects are currently under investigation by our group. 

Another approach we are currently investigating is to try to employ both the PIC and the 
fluid modeling of breakdown simultaneously in a hybrid treatment.  This is advantageous
because while the fluid modeling alone is valid for some electron energy distributions
such as Maxwellian or Druyvesteyn, it does not model the case when a secondary population of 
very high energy electrons develops in the breakdown.  We suspect that much of the 
interesting physics driving the breakdown is governed by these high energy electrons, and 
are working on using PIC particles to model them while using the fluid to model the remaining
population in the plasma.  Some preliminary work on this approach can be found in \cite{us1}.   

\vspace{1 cm}
\section{Conclusions and future work}
   We have presented here some background in breakdown theory, and two methods for numerically
modeling RF breakdown.  We have predicted that RF breakdown depends in a significant way on 
the collision frequency of electrons in the plasma with the background gas molecules, and have
demonstrated that both PIC and fluid breakdown models reproduce this dependence.  Of particular 
interest is the result that a high collision frequency can mitigate the attenuation of the 
transmitted signal. Even in the case where the plasma frequency exceeds the RF frequency, 
transmission through the plasma layer can be achieved if the density of the background gas 
is large enough to restrict the mobility of the plasma electrons.  We suggest that in this
range of parameters, ``breakdown'' does not occur because the RF signal is able to propagate 
through the medium.

   We have also reported on the results of a preliminary breakdown computation in 
a helium background gas.  We have discovered several stages in the breakdown process; the 
homogeneous heating
and ionization of the background gas, the exponential breakdown of a thin layer just at
the surface of the vacuum-helium interface, the resulting lack of transmitted power into the
interior of the gas causing a relaxation of the ionization rate everywhere but at the interface, 
the reflection of an increasing fraction of the RF power by the surface plasma layer, and the
eventual stabilization into an equilibrium with fixed density profile and ratio of reflected to 
transmitted power.  We were interested to find that once equilibrium is reached, 
almost all of the signal attenuation 
occurs at the vacuum-helium interface because the density 
is several orders of magnitude larger
there than anywhere else in the gas.  

   We experienced much difficulty in reproducing the breakdown simulation with the PIC 
model we have developed.  This is because the exponential growth in density at the 
interface with the vacuum causes the simulation to slow down a considerable amount.  The 
density growth also causes the Debye length to shrink, and at later stages in the simulation
we are no longer resolving it, resulting in spurious heating of the simulation particles.  
Finally, for very large values of the collision frequency, the random number generation 
for collisions in each cell begins to dominate the total run time, causing this method to become 
impractical.  

   In future, we will find a valid statistical approach to simulating PIC collisions
in the range of parameters requiring multiple collisions per cell per timestep.  Having learned 
that most of the interesting breakdown physics occurs in a very thin layer of gas, we will 
modify our simulation geometry in an attempt to draw a direct comparison between our two 
methods.   We will continue our PIC-fluid hybridization efforts in an attempt to simultaneously 
capture all the relevant physics. 
We will study breakdown in air rather than helium, and investigate the effects of
introducing humidity, water droplets, and dust.   We would also like to investigate pulsed 
RF signals, and contrast their impact on breakdown to the continuous wave results.  In the long
term, once we have bench-marked this machinery and compared it to the results of air breakdown
experiments currently being designed at the Air Force Research Laboratory, we can incorporate 
this technology into the virtual prototyping of future high power microwave weapons. 

\bigskip
The authors would like to thank Dr. Peter J. Turchi for
useful discussions on this work.
This research was supported in part by Air Force Office of Scientific
Research (AFOSR).

\end{document}